\documentclass[twocolumn,twocolappendix]{aastex631}
\usepackage{CJK}
\usepackage{amsmath}

\def\kms{\,{\rm km}\,{\rm s}^{-1}}

\def\feh{\hbox{[Fe/H]}}

\newcommand{\teff}{T_{\rm eff}}
\newcommand{\rv}{{\rm RV}}

\def\vsini{(V \sin{i})}
\def\logg{\log{\rm (g)}}
\def\snr{\hbox{S/N}}

\begin{document}

\begin{CJK*}{UTF8}{gbsn}

\title{Shortest period for outer orbit in compact hierarchical triple? Discovery of SB1 around V 885 Per.}

\correspondingauthor{Mikhail Kovalev}
\email{mikhail.kovalev@ynao.ac.cn}

\author[0000-0002-9975-7833]{Mikhail Yu. Kovalev (Михаил~Юрьевич~Ковалёв)}
\affiliation{International Centre of Supernovae, Yunnan Key Laboratory, Yunnan Observatories, Chinese Academy of Sciences, Kunming 650216, China}


\begin{abstract}

I present a discovery of SB1 in the LAMOST-MRS spectra of eclipsing contact binary V 885 Per. Spectroscopic orbit has period only $P=21.8$ day, which will make it shortest known outer period in compact hierarchical triple system, if this SB1 is indeed bound to the inner system. Unfortunately quality of available spectra don't allow to give definitive answer to this question,  with other orbital hierarchy like 2+2 is also possible. { Analysis of the eclipse timing variations showed a gradual increase in the orbital period of the inner eclipsing pair, as well as long-term changes that can be explained either by the light time travel effect due to an additional component or by magnetic activity. }



\end{abstract}



\section{Introduction} \label{sec:intro}
\end{CJK*}


For more than 68 years $\lambda$ Tauri was holding a record for being compact hierarchical triple (CHT) with shortest outer period $P_{out}\sim33.02$ days, but recently TIC 290061484 beat it with $P_{out}\sim24.5$ days, derived using triple eclipses detected in space based photometry \citep{2024ApJ...974...25K}. Authors suggested that CHT with even shorter period can be detected in the future. In this article I present a discovery of third star, which possibly orbits well known W UMa system V 885 Persei\footnote{discovered as variable star by Anton Valentinovich Khruslov \citep{2007PZP.....7....6K} using data from Nothern Sky Variability Survey NSVS \citep{nsvs}} with period $P_{out}\sim21.8$ days. I detected it in spectra from LAMOST medium resolution {( $\sim7500$)} survey (MRS \cite{mrs}) using similar methods to \cite{ttau}. Unfortunately here we don't have triple eclipses, like in TIC 290061484, but we have radial velocities ($\rv$) for all three components: which allows to determine a good single-lined (SB1) spectroscopic orbit for the third star, while $\rv$ measurements two components of contact system are too uncertain to confidently confirm that third star is bound to inner contact system. New set of high resolution spectroscopic observation taken at periastron passage of SB1 orbit will be very helpful for this. Also, a recent study based on astrometric data \citep{widebinary}, is suggesting that nearby star {\it Gaia DR3} 447429768548995456 (G4474 hereafter) visible at $\sim3.1$\arcsec, forms a wide system with V 885 Per, making it high multiple system.

\section{Methods and results} 
\label{sec:methods} 
\subsection{Spectroscopy}
\begin{figure*}
    \centering
    \includegraphics[width=0.95\textwidth]{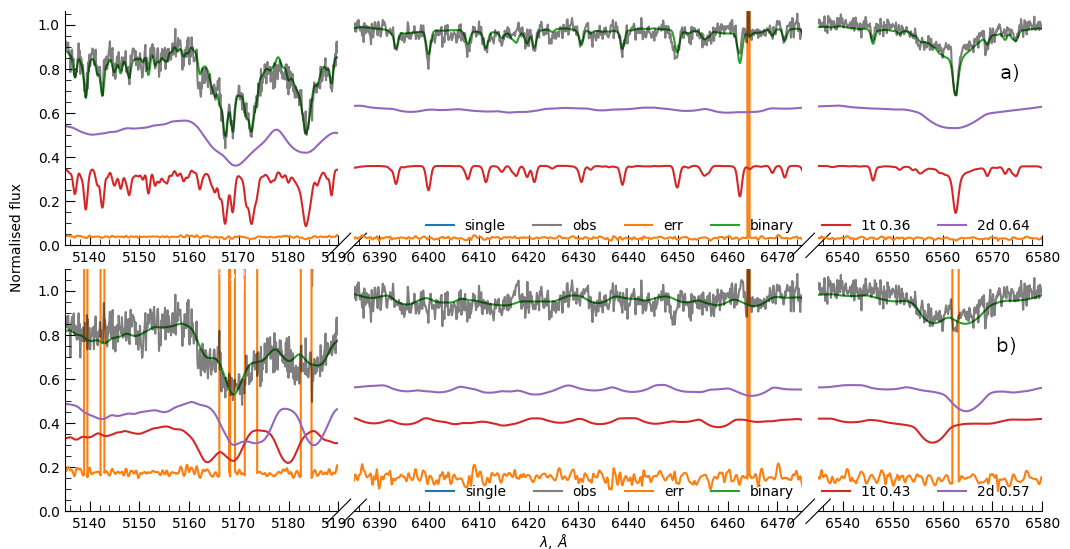}\\
    \includegraphics[width=0.45\textwidth]{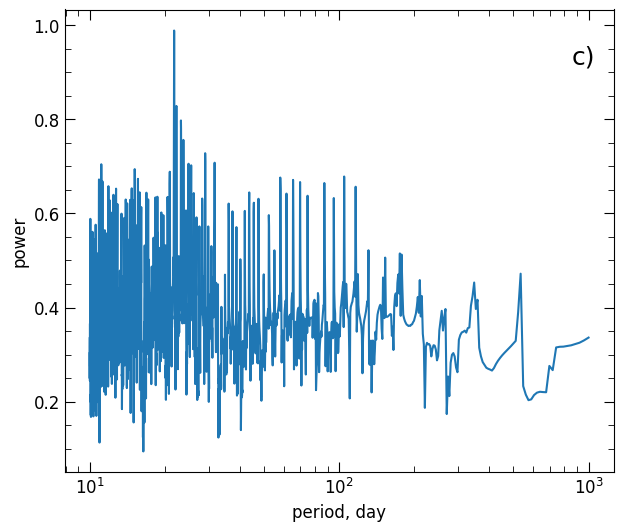}
    \includegraphics[width=0.45\textwidth]{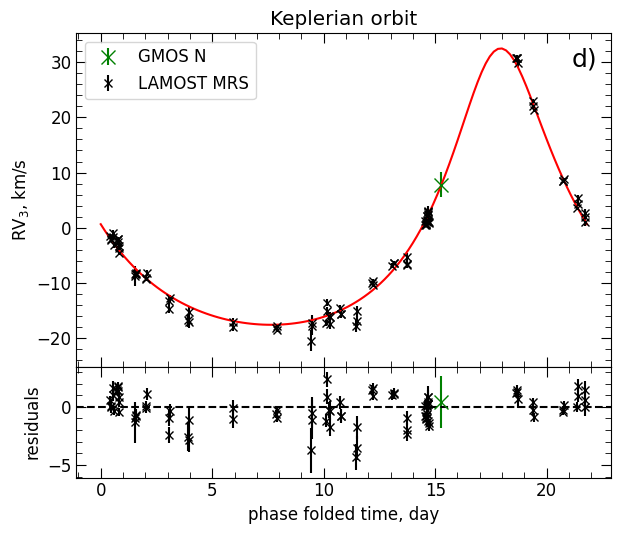}
    \caption{Panels a) and b): spectral fitting example in two iteration (inner binary phase $\phi=0.71$). Panel c): GLS periodogram, Panel d): Keplerian orbit fit for star B.  } 
    \label{fig:spexampl}
\end{figure*}

LAMOST MRS provides spectra of V 885 Per under designation J030621.69+544702.2 at \url{lamost.org/dr11}.
All available 89 spectra were observed using 20 minute exposure time and were analyzed using binary spectral model with two iterations, similarly to \cite{ttau} (check it for detailed description). They span interval of 1147 days (BMJD=$58801.6:59948.5$ d) and have mean $\snr_{\rm blue,red}=10,20\,{\rm pix}^{-1}$ (full range is $\snr_{\rm blue,red}=4:28 , 11:57~{\rm pix}^{-1}$). Both spectral arms were fitted simultaneously, with wavelength scale corrected to possible zero point shifts, following \cite{zb_rv}. Resulting fit after first iteration shows composite structure of the spectrum, see Figure~\ref{fig:spexampl}: one narrow-lined component ($\vsini=1\pm2~\kms$) and another component has very broad spectral lines (variable with time $\vsini=150:300~\kms$). The narrow line component contributes from one third to one half of total light, which is consistent with light curve (LC) based on ASAS-SN $g$ band photometry, see following Section~\ref{sec:wd}. Following ``Matryoshka" method developed in \cite{ttau}, I subtracted it's contribution from the spectrum and applied binary model to fit residual spectrum. Error spectrum was increased according to subtracted fraction. To avoid binary model from the fitting of possible subtraction artifacts I set very large errors for wavelength regions where scaled subtracted model was below 0.2. Results match the expectation for contact system: $\rv$s follow circular orbit with same period as LC, with $\rv$ amplitude ratio around two. Overall these $\rv$ measurements have low precision, due to very broad lines and low $\snr$ of residual spectra, but this is compensated by large number statistics, so they can be used to get basic understanding on mass ratio and systemic velocity of inner system. More precise measurements are possible with spectral disentangling, see Section~\ref{sec:korel}. I computed mean and standard deviation of all fitted parameters using all spectra for all three spectral components in Table~\ref{tab:1}. 

\begin{table*}
    \centering
    \begin{tabular}{c|cccc}
        Parameter/Component & Aa (1)& Ab (2) & B (3) & G4474\\
        \hline
        Gaia DR3 (a)\\
        $\varpi$, mas & \multicolumn{2}{c} {2.699$\pm$0.267}& & 2.935$\pm$0.078\\
        $\mu_\alpha\cos{\delta}$, mas/yr & \multicolumn{2}{c} {16.420$\pm$0.236}& & 17.116$\pm$0.070\\
        $\mu_\delta$, mas/yr & \multicolumn{2}{c} {-12.273$\pm$0.270}& & -12.008$\pm$0.084\\
        RUWE & \multicolumn{2}{c}{19.746} & & 0.904\\
        $G$, mag & \multicolumn{2}{c}{13.229$\pm$0.007} & & $17.164\pm0.003$\\
        $BP$, mag & \multicolumn{2}{c}{13.98$\pm$0.02} & & $18.16\pm0.02$\\
        $RP$, mag & \multicolumn{2}{c}{12.36$\pm$0.02} & & $15.79\pm0.02$\\
        \hline
        Ephemeris (b)\\
        $P$, day &\multicolumn{2}{c} {0.3038707}& \\
        $t_0$(*), (HJD-240000.5) day &\multicolumn{2}{c} {57936.0165$\pm$0.0007}& \\
        \hline
        Spectroscopic analysis\\
        $\teff$, K & 5019$\pm$98 & 5204$\pm$127 & 5139$\pm$116\\
        $\logg$, cgs &  3.9$\pm$0.3 &4.1$\pm$0.3 & 3.9$\pm$0.3\\
        $\feh$, dex & -0.1$\pm$0.1& -0.1$\pm$0.1& -0.1$\pm$0.1\\
        $\vsini,\,\kms$ &  154$\pm$30 &  155$\pm$40 & 1$\pm$2 \\
        \hline
        SB1 orbit\\
        $P$, day & & & 21.811$\pm$0.001\\
        $t_p$, (BMJD) day & & & 58819.06$\pm$0.09\\
        $e$ & & & 0.39$\pm$0.01\\
        $\omega^\circ$ & & & 354.3$\pm$1.5\\
        $K, \,\kms$ & & & 25.13$\pm$0.43\\
        $\gamma, \,\kms$ & & & -2.26$\pm$0.17\\
        $f(M),\,M_\odot$ & & & $0.0278\pm0.0015$\\
        \hline
        Wilson-Devinney model \\
        $\teff$ & 5000 & 5200\\
        $i^\circ$ &\multicolumn{2}{c} {75$\pm$1}& \\
        $\Omega$ &\multicolumn{2}{c} {3.04$\pm$0.01}& \\
        $q$ &\multicolumn{2}{c} {0.59$\pm$0.02}& \\
        $a,\,R_\odot$ &\multicolumn{2}{c} {2.18$\pm$0.04}& \\
        $L_1$ &\multicolumn{2}{c} {4.86$\pm$0.03}& \\
        $L_3$, \%&\multicolumn{2}{c} {34.3$\pm$0.4}& \\
        $R,\,R_\odot$ & 0.93 & 0.73 & \\
        $M,\,M_\odot$ & 0.95 & 0.56 & \\
        $\log{L/L_\odot}$ & -0.30& -0.46 & \\
        \hline
        SED fit\\
        $R,\,R_\odot$ & 0.93(fixed) & 0.73(fixed) & 0.98\\
        $\teff$, K & 5005 & 5143 & 5050 \\
        $d$, pc &\multicolumn{3}{c}{335}\\
        $A_V$, mag & \multicolumn{3}{c}{1.40}\\
    \end{tabular}
    \caption{V 885 Per parameters.(*) I add half period for consistency with W-D model. (a) \cite{gaia3}, (b) \cite{asassn_nn_12k}. RUWE -renormalized unit weighted error.}
    \label{tab:1}
\end{table*}

\par
The good quality of $\rv$ from narrow lined component allows direct determination of SB1 orbit. Period $P=21.811\pm0.001$ days was found using Generalised Lomb Scargle code \citep{gls}, see periodogram in Figure~\ref{fig:spexampl}, with further refinement of solution using $\chi^2$. All parameters are collected in Table~\ref{tab:1}. Orbit has moderate eccentricity $e_B=0.39$, with some measurement taken near periastron passage, see Figure~\ref{fig:spexampl} panel (e). Unfortunately contact system has eclipse at this moment, so other two components have $\rv$ of low quality, see Appendix~\ref{sec:peri}.  I found that V 885 Per was observed by Gemini North on MJD=54781.5 d using grating R831 and exposure 300 seconds. Resolution of this spectrum is only $\sim4400$, but it allow to measure $\rv=7.8\pm2.2~\kms$ for narrow lined component (labeled as GMOS N in panel (d) of Figure~\ref{fig:spexampl}), which perfectly match orbital solution. 


\subsection{W-D model}
\label{sec:wd}
\begin{figure}
    \centering
    \includegraphics[width=\columnwidth]{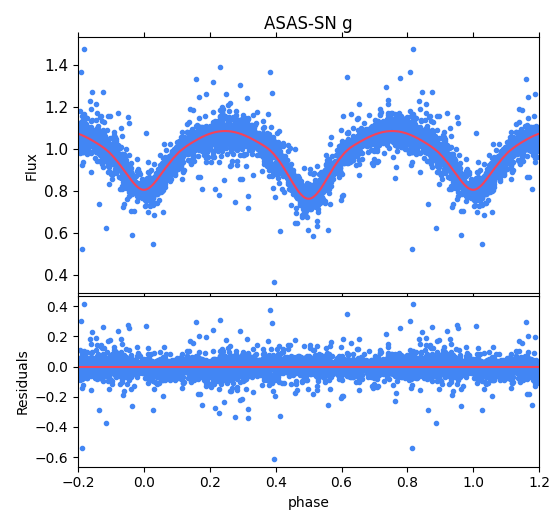}\\
    \includegraphics[width=\columnwidth]{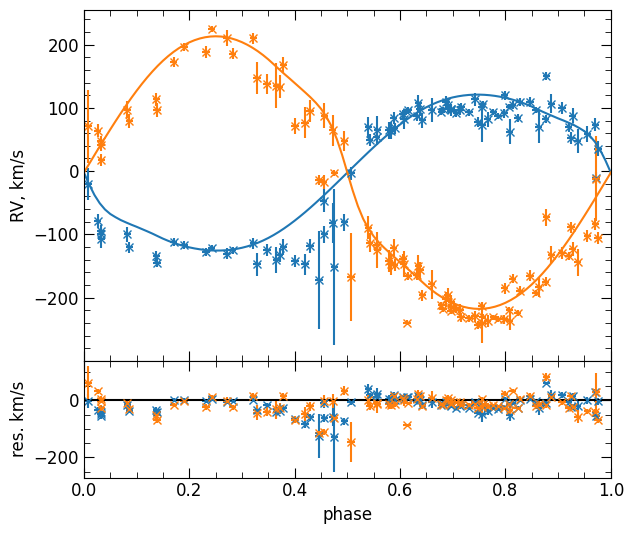}
    \caption{W-D solution for ASAS-SN $g$ LC and ``Matryoshka" based RVs.}
    \label{fig:w-d}
\end{figure}

Light curve model can provide useful test for results derived solely from the spectra. V 885 Per was observed by multiple photometric surveys, however I decided to choose LC from ASAS-SN \citep{asassnv,asassn2023} $g$ band because it is close to LAMOST-MRS and has many observations (3198) overlapping with times of spectra (HJD$=2458311:2460765$ d). More precise photometry from {\sc TESS} \citep{tess} is available in several sectors, but it covers infrared region which isn't overlapping with LAMOST-MRS.
I use the latest version of Wilson-Devinney code ({\sc W-D}) \citep{wd71,wilson79,wilson14} with {\sc PYWD2015} \citep{pywd} user interface to fit LC and RV for a inner contact system. Fitted parameters for differential correction program include dimensionless potential $\Omega$, primary bandpass luminosity $L_1$ and third light contribution $L_3$ for $g$ band, semimajor axis $a_A$, mass ratio $q_A$ and inclination angle $i_A$. $\teff$ were fixed using values from spectra. Systemic velocity was fixed to $\gamma$ from SB1 solution. Gravity darkening coefficients were set to 0.32 for both components, bolometric albedos were kept at default values of 0.5. Limb darkening coefficients were computed internally by W-D, based on atmospheric parameters of components. Resulting fit is shown on Figure~\ref{fig:w-d}, with parameters collected in Table~\ref{tab:1}. LC is fitted well, while RV shows good match overall, although there are several large outliers near eclipses, where spectra have lower $\snr$. 
 Recently \cite{asassn_nn_12k} analysed this LC with Neural network based algorithm. Without RV measurements they got much smaller mass ratio $q=0.216\pm0.004$ and much higher third light contribution $L_3=0.594\pm0.007$. 

\subsection{SED fit}
\begin{figure}
    \includegraphics[width=\columnwidth]{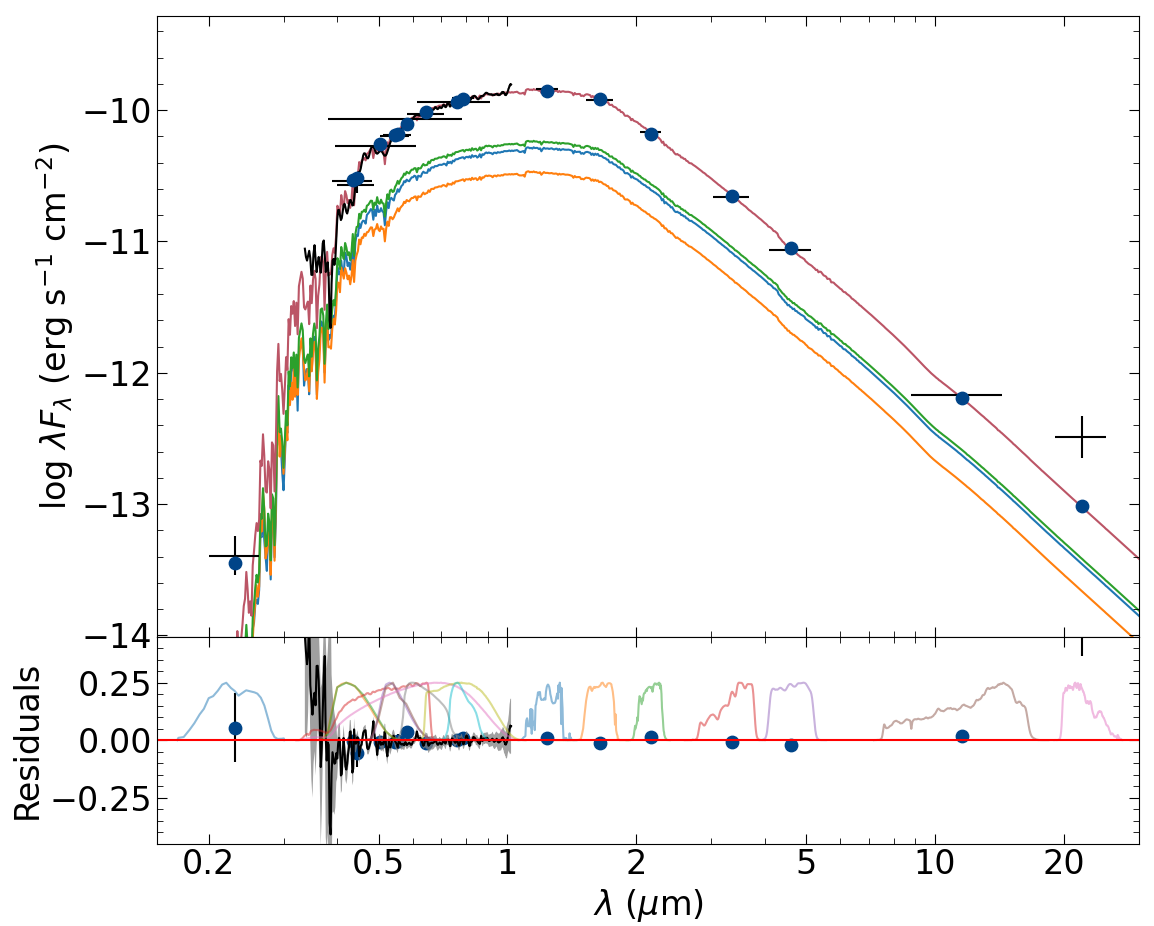}
    \caption{SED fit for V 885 Per. Red line shows combined SED for all three components. Black line is a Gaia XP spectrum. Blue and orange lines are primary and secondary components of inner eclipsing binary. Green line is Star B. }
    \label{fig:spexampl2}
\end{figure}

I use SEDFit code \citep{sedfit} to fit Spectral Energy Distribution (SED). Archival photometry was downloaded from VizieR ($Johnson~B,V$, $Cousins~U,B,V,R,I$, GALEX $NUV$, Gaia DR3 $G,BP,RP$, 2MASS $J,H,K_s$, WISE $W1,W2,W3,W4$).  For GALEX and 2MASS times of observations correspond to phases $\phi=0.62, ~0.42$ of inner system, so both components contribute to SED. Gaia and WISE are averages of many observations during surveys, so photometry covers many phases of LC, including minima. Therefore I checked LC and confirmed that mean values corresponds to LC phases when both components are visible. All these data were fitted by combined flux of three components. $W4$ measurement is reported as higher limit, so it wasn't used during the fit. SEDFit allows fitting of Gaia XP low resolution spectrum, however I don't fit it, but use it for verification.
Parameters of contact inner pair were constrained using W-D model, while spectroscopic solution puts limits for parameters like $\teff$, with surface gravity and metallicity were fixed to $\logg=4.43, \feh=-0.1$ dex for all three components.
I use spectral models from BT-Settl \citep{btsettl}. Parallax from Gaia DR3 \citep{gaia3} was used to initialize distance value. Unfortunately current version of SEDFit doesn't estimate uncertainties.  
Resulting fits are shown in top panel of Figure~\ref{fig:spexampl2}, while lower panel is showing residuals with schematic image of the transmission function for used filters. Generally agreement is very good. Gaia XP spectrum also agrees with model, however we can see slight mismatch in the blue part. 
The best fit parameters are: $d=335$ pc, $A_V=1.40$ mag, $R_{B}=0.98~R_\odot$, ${\teff}_{Aa,Ab,B}=5000, 5143, 5050$ K, see Table~\ref{tab:1}.
Components Aa and B are very similar, but component Ab is slightly smaller and hotter. 

\subsection{Eclipse Timing variation analysis}
\label{sec:etv}
\begin{figure*}
    \includegraphics[width=\textwidth]{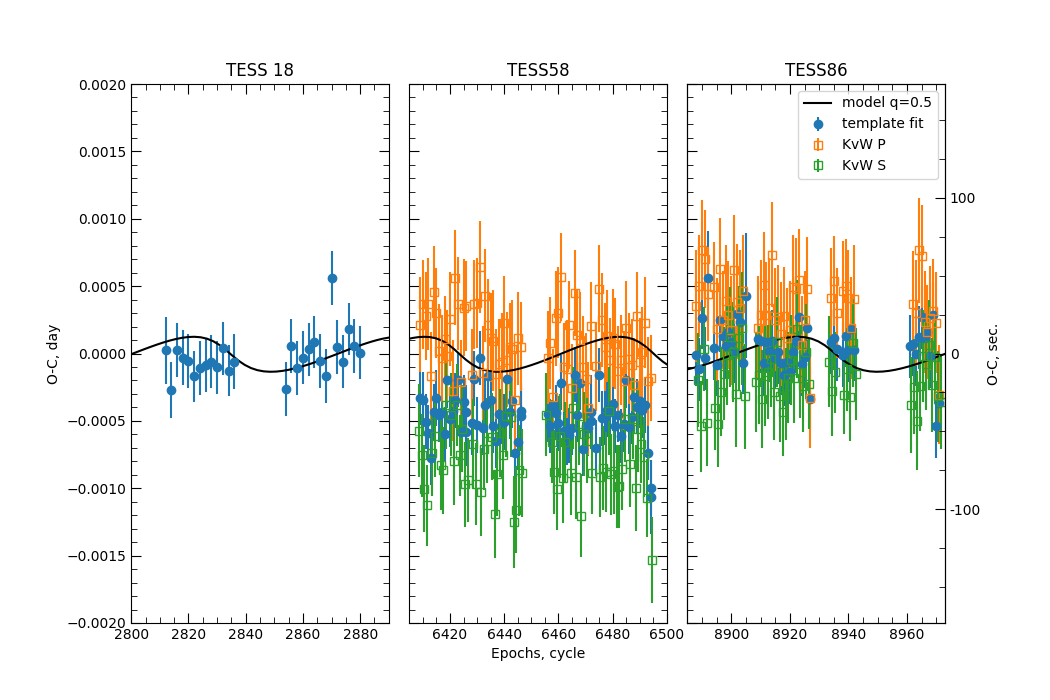}
    \caption{Comparison of ETV measurements with template fitting and KvW methods for LC from three TESS sectors. Expected LTTE computed assuming mass ratio $M_B/M_A=0.5$ is shown as a black line. } 
    \label{fig:kvw}
\end{figure*}

Eclipse timing variation (ETV) can provide useful information about motion of inner system \citep{2022Galax..10....9B}. I downloaded all available TESS photometry from MAST portal\footnote{\url{doi.org/10.17909/3em1-sh77}}, using TESS-Gaia light curve package \citep{tglc}. There are three sectors available: 18 with cadence 1800 s, 58 and 86 with cadence 200 s. Calibrated Point Spread Function (PSF) photometry was selected with quality flag equals to zero. Traditional Kwee-van Voerden (KvW) method \citep{kvw} can be applied only for last two sectors, because they contain enough datapoints around each minima. I used KvW implementation from \cite{kvwnew}. First sector can be analysed using template fitting method similarly to \cite{ttau} when I use all datapoints phased using two period intervals. For other two sectors this method can use measurements in one period time intervals. This method requires precomputed templates for each sectors, which were computed using W-D, see Appendix~\ref{sec:tess}. Median time point in the interval was used as an epoch of the measurement. This template matching technique allows to take advantage of the characteristic shape of W UMa LC \citep{no2kvw} and not require equidistant time sampling like KvW method. I compare these two methods for two LC from TESS sectors 58 and 86 in Figure~\ref{fig:kvw}. Expected light time travel effect (LTTE) due to motion of inner system's barycenter, computed using SB1 orbit and mass ratio $(M_B/M_A=0.5)$ is also shown. It is clear that using precise TESS data one can't see fast changing reflex motion caused by third star, when the third star will be less massive than inner system. \cite{2022Galax..10....9B} stated that no detection of LTTE is possible if $\mathcal{A_{\rm LTTE}}_A\leq50$ s. Tiny variation caused by LTTE is completely lost in the uncertainties of the measurements. I did series of simulations using synthetic LC generated at exact times of TESS observations with artificially included LTTE, and found that LTTE can be confidently detected only if $M_B/M_A\geq1.0$, see Appendix~\ref{sec:tess} for details.  

It is clear, that O-C values for TESS 58 are smaller than for two other sectors. It worth to explore ETV and times of minima from O-C gateway database\footnote{\url{var.astro.cz}} \citep{omc_gateway} over large interval of time, which can reveal LTTE caused by additional long period components or period change. Additional light curves were provided by ASAS-SN\citep{asassnv},  SuperWASP\citep{superwasp}, ZTF\citep{ztf_main} surveys. For several minima O-C gateway provided full LC, which can also be analysed. Unfortunately NSVS data, used to discover this system, are not available anymore, but I did a successful attempt to recover it using plot from \cite{2007PZP.....7....6K}, see Appendix~\ref{sec:ai}. 
Times of minima were computed using W-D from full LC, see Appendix~\ref{sec:ai}, and subtracted {\bf ephemeris} from Table~\ref{tab:1}. The O-C curve was modeled using parabolic plus periodic variation, in a form of Keplerian orbit, which should take into account the LTTE effect. 

\begin{align}
    O-C=\Delta t_{min} + \Delta P~E + \beta~E^2+W(p_{\rm Kep},E),
\end{align}
where $\Delta t_{min},~\Delta P$ are corrections for linear ephemeris, $\beta$ is a long-term change of the period and $W$ is a periodic variation, where $p_{\rm Kep}=\mathcal{A},~P,~t_{p},~e,~\omega$ are parameters of the Keplerian orbit. I used a built-in Keplerian model from PyAstronomy\citep{pyasl}, which allows the computation of position and velocity along the line of sight for a two-body problem.

\begin{figure*}
    \includegraphics[width=\textwidth]{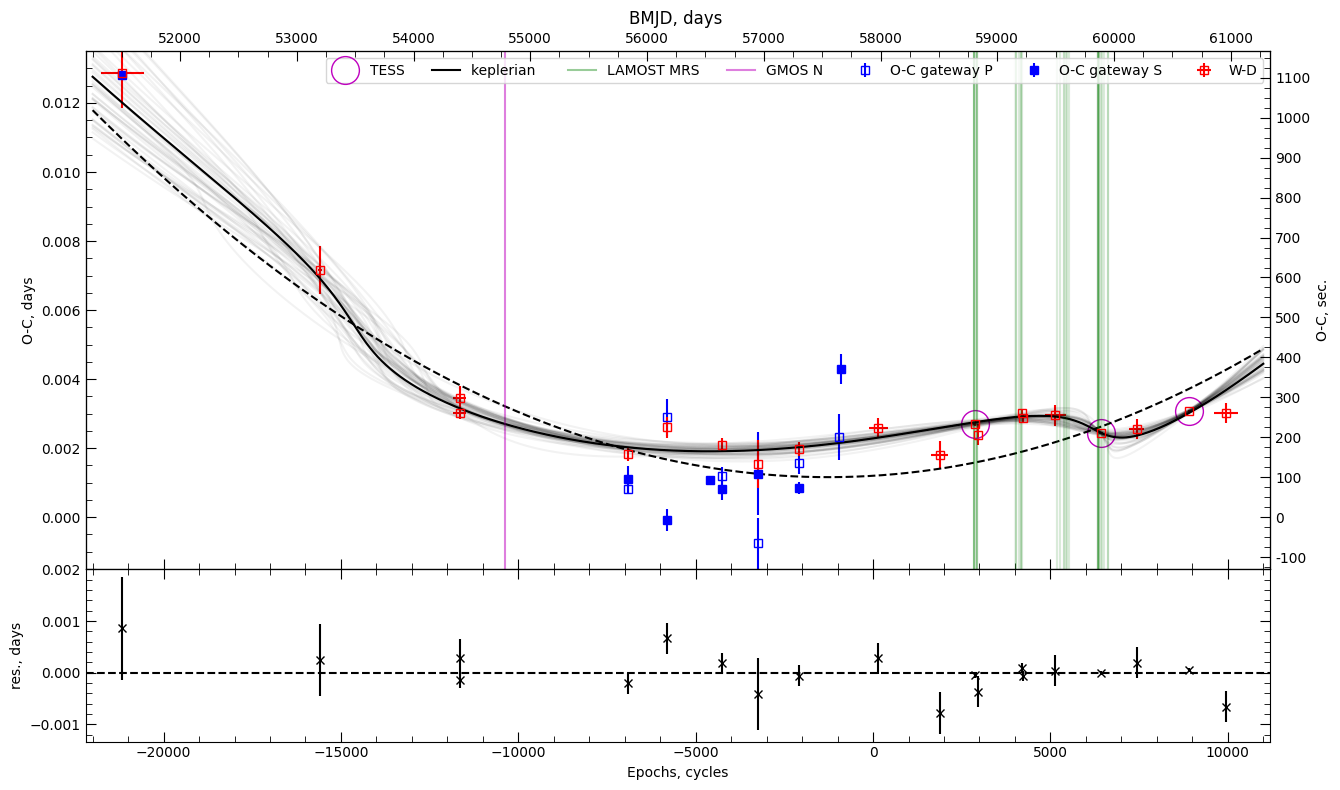}
    \caption{O-C curve fit using parabola (dashed line) plus Keplerian orbit (solid line). Gray lines shows 100 randomly chosen models from {\sc EMCEE} sample. Horizontal axes are shown with units in cycles (bottom) and BMJD (top) for convenience. Vertical lines indicate times of spectral observations. Residuals are shown in the bottom panel. TESS datapoints are highlighted by open circles. Datapoints corresponding to O-C gateway provided minima times are shown, but not used during fit.}
    \label{fig:omc}
\end{figure*}

I sample solutions with {\sc EMCEE} \citep{mcmc} using 48 walkers and 30000 iterations (first 25000 were used as {\bf a burning} stage).  We present the resulting O-C curve fit in Figure~\ref{fig:omc}. It is clear that my measurements computed using LC agree with model significantly better than O-C gateway values. Note that very precise TESS datapoints work as anchors.

Table~\ref{tab:o-c} lists fitted parameters. I compute estimates and their uncertainties as a median and standard deviation of {\sc EMCEE} sampling. $\beta\geq0$, thus the orbital period of the contact binary increases at a very slow rate of $dP/dt=3.0\cdot10^{-8}~{\rm d~yr^{-1}}$ (a second in 400 years).
The updated ephemeris is:
\begin{align}
t_{min}(\rm BMJD)= 57936.0179(3)+0.30387076(1) E
\label{eq:efimer1}
\end{align}


\begin{table}
    \centering
    \caption{O-C curve fit. Estimates and errors are from median and standard deviation of {\sc EMCEE} sampling.}
    \begin{tabular}{lc}
Parameter & value \\
\hline
$\Delta t_{min}$, day &  0.0012$\pm$0.0003 \\
$\Delta P$, $10^{-8}$ day cycle$^{-1}$&  6.3$\pm$1.4\\
$\beta$, $10^{-10}$ day cycle$^{-2}$  &  0.25$\pm$0.02\\
$\mathcal{A}$, day &    0.00130$\pm$0.00033 \\
$P$, day &  6341$\pm$617  \\
$t_{p}$, (BMJD) day  & 59861$\pm$145 \\
$e$ &  0.70$\pm$0.13\\
${\omega} ^\circ$  & 195$\pm$13\\
computed:\\
$K,~\kms$& 0.66$\pm$0.86\\
$f(M)=\frac{(M_C \sin{i_C})^3}{M^2_{tot}},~10^{-4}\cdot M_\odot$ & $0.9\pm5.4$\\

    \end{tabular}
    \label{tab:o-c}
\end{table}

{ The LTTE offers a plausible interpretation of the long term changes in O-C, but the resulting Keplerian orbit is highly uncertain. As an alternative, the period modulation observed in V 885 Per may arise from magnetic activity via the Applegate mechanism \citep{Applegate_1992ApJ...385..621A}, a scenario that is qualitatively supported by the time-varying O'Connell effect seen in many light-curve datasets (see Appendix~\ref{sec:ai}).  }


\subsection{Spectral disentangling}
\label{sec:korel}

Spectral disentangling technique allows one to separate individual spectral components and even avoid the direct estimation of radial velocities \citep{korelman}. I used {\sc KOREL} code \citep{korel}, available through virtual observatory (VO-KOREL) service \citep{vokorel}. It works in Fourier space and can handle up to five spectral components with different hierarchical orbit configurations. All input spectra of V 885 Per were cleaned from cosmic ray contamination, while contaminated regions were interpolated using values from nearby pixels. Spectral region around Mg triplet was used to make input file with 2048 datapoints for each of 89 spectra. {\sc KOREL} allows to fit for relative line-strengths of each spectral component in every input spectrum, therefore all spectra were used even ones from eclipses. Radial velocities are also estimated by {\sc KOREL} once it have disentangled spectral components.
\par
\begin{figure*}
    \centering
    \includegraphics[width=0.9\linewidth]{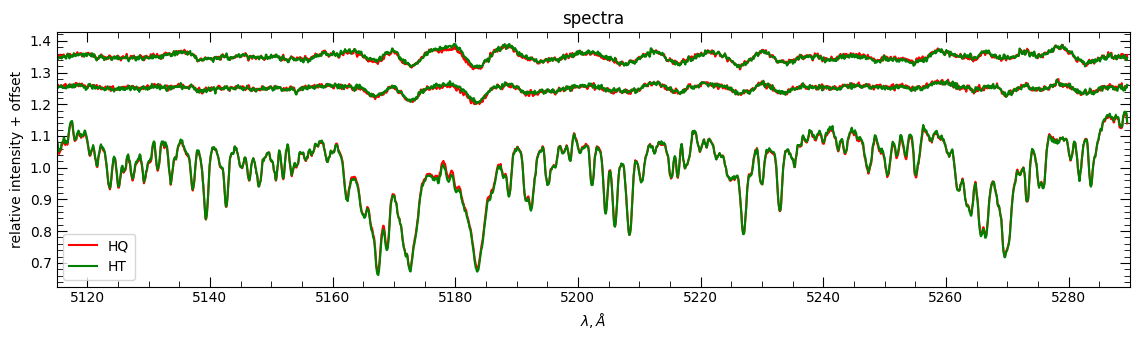}\\
    \includegraphics[width=0.45\linewidth]{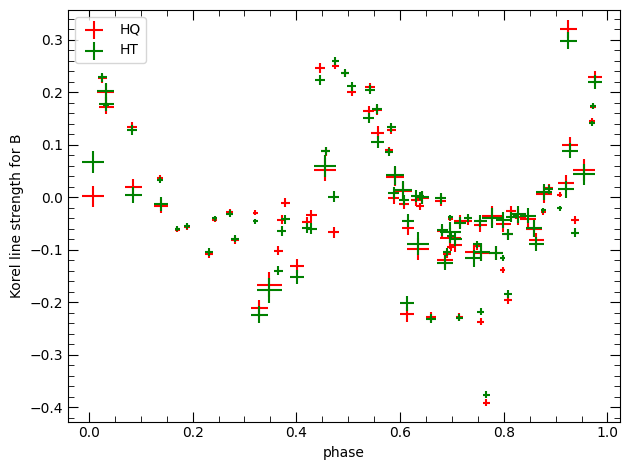}
    \includegraphics[width=0.45\linewidth]{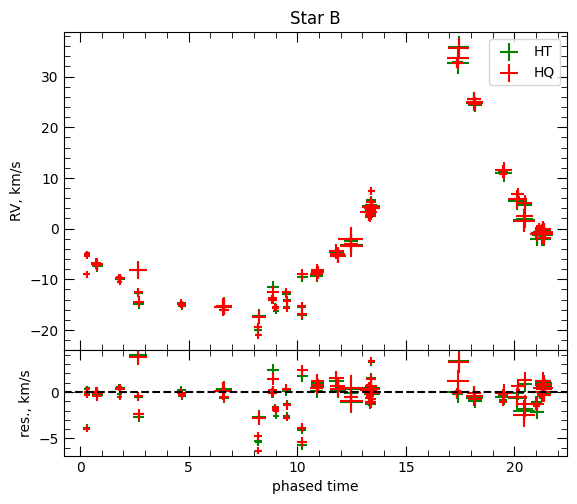}\\
    \includegraphics[width=0.45\linewidth]{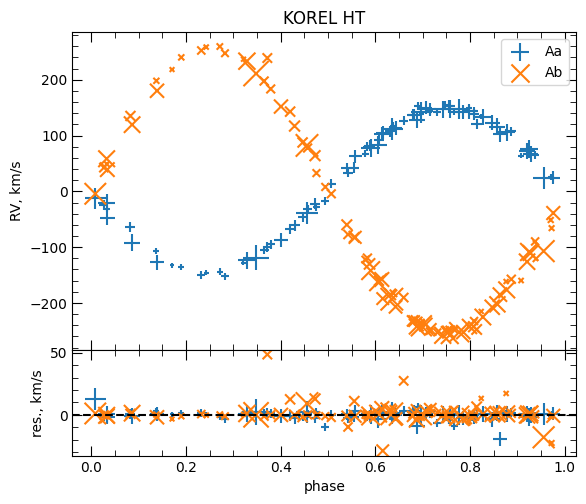}
    \includegraphics[width=0.45\linewidth]{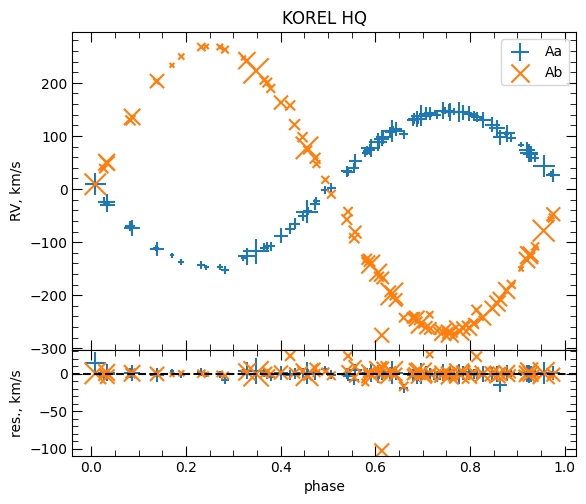}\\
    \caption{{\sc KOREL} output for two configurations: three disentangled components (top), line-strength of Star B vs phase of eslipsing pair (middle left), RV for star B (middle right), RV for components in contact pair (bottom panels). Size of data points is proportional to $\snr$ of input spectra. }
    \label{fig:korel}
\end{figure*}

At first I tried hierarchical triple (HT) configuration (2+1): circular orbit of inner pair Aa+Ab and star B on elliptical orbit with period 21.8 day around it. Periods were fixed, but $e_B, \omega_B, t_{p}$ and radial velocity amplitudes for both orbits were set as free parameters. All three spectral components were separated successfully: lines in components of inner contact system are significantly broadened in comparison with star B, which supports  spectral analysis results. Derived spectra of components and radial velocities are shown in Figure~\ref{fig:korel} with orbital parameters listed in Table~\ref{tab:korel}. Line strengths of star B are consistent with minima of LC. Inner pair mass ratio is in good agreement with W-D solution. Mass ratio $M_B/M_A=1/1.82=0.55$ supports absence of LTTE with period of 21.8 day in ETV analysis of TESS data, see Appendix~\ref{sec:tess}.

\setlength{\tabcolsep}{8pt}
\begin{table*}
    \centering
    \begin{tabular}{c|cc|cc}
    \hline
    Parameter & \multicolumn{2}{c}{HT}  & \multicolumn{2}{c}{HQ} \\
    \hline
    orbit & A & B  & A & B  \\
      $P$ (fixed), day   &  0.30387076 & 21.811 &0.30387076 & 21.811 \\
    $t_p$, day   & 57936.016$\pm$0.001 & 58819.071$\pm$0.003 & 57936.017$\pm$0.001 & 58819.054$\pm$0.005 \\
    $e$          & 0.0 (fixed)& 0.390$\pm$0.001 & 0.0 (fixed) & 0.390$\pm$0.001 \\
    $\omega^\circ$ & 90 (fixed)& 354.9$\pm$0.1 & 90 (fixed) & 354.2$\pm$0.4 \\
    $K_1,~\kms$ & 149.4$\pm$0.3 & 25.45$\pm$0.05 & 146.7$\pm$1.7 &25.34$\pm$0.17\\
    $q=K_1/K_2$ &  0.584$\pm$0.001 & 1.82$\pm$0.01 & 0.544$\pm$0.007 &  \\
    $K_2,~\kms$ & 256.0 & 14.0 & 269.7 & \\
    RMS(RV), $\kms$ & 2.16,6.70  & 0.97   &  2.41,13.60 & 0.97\\
    \hline
    \end{tabular}
    \caption{Orbital parameters from KOREL for two orbital configurations.}
    \label{tab:korel}
\end{table*}

\par
Additionally, I investigated a hierarchical quadruple (HQ) configuration of the (2+2) type, consisting of an inner contact binary A and a second stellar pair B, where the primary is accompanied by a faint (or possibly compact) companion on an elliptical orbit with a period of 21.8 days; both subsystems orbit a common center of mass with large period ($P_{AB}>>21.8$ day). Fit was initialized as before, with zero RV amplitudes for outer long period orbit, which means star B not affecting inner pair. Again all three spectral components were separated successfully, see Figure~\ref{fig:korel}, with almost identical spectral profiles as HT, Also inner pair have slightly smaller mass ratio.
\par
RV curves look very similar for both HT and HQ configurations, however HQ showing more outliers for components Aa and Ab. RMS statistics of RV is better for HT configuration: $RMS(RV)_{Aa,Ab,B}=2.16, 6.70,0.97$, HQ: $RMS(RV)_{Aa,Ab,B}=2.41, 13.60,0.97$, therefore first configuration should be preferred.

\section{Discussion}

\begin{figure*}
    \centering
    \includegraphics[width=\linewidth]{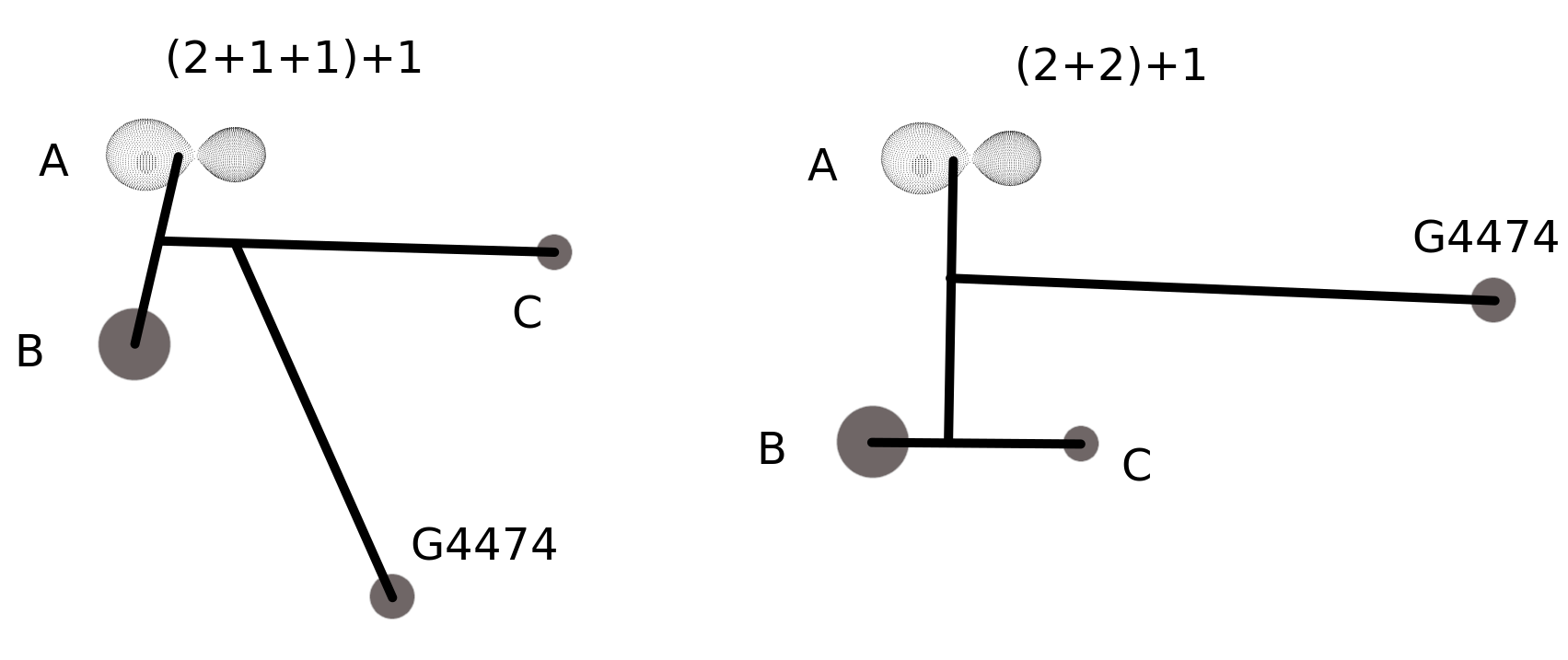}
    \caption{Schematic image for possible hierarchical configurations of V 885 Per.}
    \label{fig:cartoon}
\end{figure*}
Photometry shows W UMa type LC with period $P_A=0.30387$ day, which is supported by spectroscopic observations, where one can identify three distinct stellar components: two with broad lines and one narrow lined with SB1 period of 21.8 days. ETV analysis unable to support this period, but it suggests that barycenter of eclipsing pair moves with longer period of $\sim6300$ days through LTTE, although this detection is not very clear.
To explain these results one can assume hierarchical quadruple system in two competing configurations: CHT+1 (2+1+1 configuration, like CW Cas in \cite{cwcas}) and HQ (2+2). If we take into account wide companion G4474 found by \cite{widebinary} it is even quintuple, see Figure~\ref{fig:cartoon}. For the first one it is reasonable for inner binary to have same systematic velocity as SB1 solution, while for the second one systemic velocities of inner binary and SB1 should change with time in opposite phase. 
Both spectra and ETV allow us to explore properties of this multiple system in projection on the line of sight, and one always need to do some assumptions on inclinations (except inner eclipsing pair where inclination is constrained).  
\subsection{CHT+1}
Results from analysis with {\sc KOREL} suggest that star B with SB1 period of $P_B=21.8$ day is likely bound to eclipsing system V 885 Per. Unfortunately quality of available spectra don't allow us to { clearly} see reflex motion of { inner pair} barycentre: spectral lines of components in the contact system are too broad for this, plus spectrum near periastron passage was taken during the eclipse of inner pair, see Appendix~\ref{sec:peri}. Future high resolution observations with large telescopes, scheduled to observe periastron passage outside the eclipse will be very useful to confirm it. In principle, since inner system is already in contact and have very short orbital period ($P_A=0.30387\, {\rm d}\sim7$ hrs, $P_B/P_A\sim72$) dynamical effects should be weak here \citep{2014AJ....147...45C}, unlike what we can see for TIC 290061484. Plus amplitude of LTTE($\mathcal{A_{\rm LTTE}}_A\leq22$ s, assuming contact pair is more massive than star B) seems to be below current detection limit of 50 s \citep{2022Galax..10....9B}, unlike J04+25 where LTTE was confidently detected \citep{ttau}. Unlike TIC 290061484 where times of minima can be easily measured from {\sc TESS} data due to ``sharp" eclipses of detached inner system, here LC minima are less defined and can be affected by the stellar activity (i.e O'Connell effect \citep{oconnel,oconnel1,oconnel2}, see Appendix~\ref{sec:ai}).  
\par
If we assume that star B have same mass as component Aa, following SED results, we will get $(a_A+a_B)\sin{i_B}\sim24 R_\odot$. Taking into account total mass ($M_A=1.5~M_\odot$) of eclipsing pair from W-D model and a Third Kepler's law we get $(a_A+a_B)\sim90 R_\odot \sim 0.41$ a.u. and $i_B\sim10^\circ$, so inner and outer orbits can be almost perpendicular to each other $i_A+i_B\sim87^\circ$.  Relatively small value of SB1 mass function  $f(M)=0.0278\pm0.0015~M_\odot$ also supports small $i_B$. Taking into account distance from SED fit sky projection of outer orbit is $\sim1.25$ mas, which comparable to parallax of V 885 Per $\varpi=2.699\pm0.267$ mas from Gaia DR3. Large value of renormalised unit weight error RUWE=19.746 indicates that single star astrometric solution is not reliable. $Gaia$ is unable to resolve outer system, but can see motion of the photocenter, although eclipses of inner system will cause additional ``shaking" to it.  


\subsection{HQ}
Alternative explanation can be that star B orbits another star.  At distance $\sim3.1$\arcsec from V 885 Per there is star G4474, which seems to form a wide binary with V 885 Per, based to analysis of Gaia astrometry by \cite{widebinary}. LAMOST MRS fiber diameter is $~3.3$\arcsec so G4474 can be observed together with our target. According to \cite{widebinary}, chance-alignment probability is only $\sim10^{-4}$ and estimated projected separation $a\sim1169$ a.u. with an eccentricity $e=0.68^{0.89}_{0.42}$\footnote{ given with upper and lower limits of 68\% credible interval}. This wide binary will show some LTTE for V 885 Per, although period should be very long $P > 1000$ years based on distance and sky projected separation, thus it is unlikely to be responsible for motion with $P\sim6300$ days detected in ETV. \cite{bjdist21} distance for this star $d=334_{326}^{342}$ pc perfectly matches our distance estimate from SED for V 885 Per. However {\it Gaia DR3} photometry of G4474 strongly disfavors hypothesis of Star B orbiting it: $G=17.171\pm0.003$ mag, $BP=18.16\pm0.07$ $RP=15.79\pm0.02$ mag, because in this case G4474 should be much brighter, roughly 40\% of V 885 Per ($G\sim14.2$ mag). 
\par 
SB1 mass function $f(M)=\frac{(M_C \sin{i_C})^3}{(M_B+M_C)^2}=0.0278\pm0.0015~M_\odot$ can provide us $M_C$. Adopting $M_B=0.8~M_\odot$ and $i=75^\circ$ same as eclipsing pair A, component C should be very light $M_C=0.34$ and faint. It's spectral lines will be well below noise level of LAMOST-MRS spectra. In principle, if it is main sequence red dwarf it should reveal itself as infrared excess in SED, but nothing is visible there, see Figure~\ref{fig:spexampl2}.  Total mass $M_B+M_C\sim1.1~M_\odot$ taken together with very small value of mass function from ETV analysis $f(M)=0.9\pm5.4\cdot10^{-4}~M_\odot $ indicates that long-period orbit has very small inclination to sky plane, which should be, in principle, easily detected by Gaia.
\par


\section{Conclusions}
I present a possible discovery of shortest outer period in compact hierarchical triple system. If confirmed it will be rare discovery of CHT in the region of ``W UMa dessert", as such systems are very difficult target to be detected using photometric observations.  SB1 solution is consistent with results from spectral disentangling. Unfortunately quality of currently available spectra don't allow me to detect expected reflex motion for barycenter of the inner contact system. Hopefully future high-resolution spectroscopic observations, scheduled using SB1 orbit to observe system at periastron passage, will support this findings. In principle even not very high-resolution spectra will be sufficient to measure reflex motion of barycenter of $\gamma_A=-15~\kms$, if it will be taken around phase of inner system $\phi=0.25$ or $\phi=0.75$, when $\rv_{Aa,Ab}$ change slowly. 
\par
 Very large RUWE=19 gives hope that future data from Gaia DR4 (expected on 2026-12-02) will be very useful for analysis of V 885 Per, since the detection of astrometric orbit consistent with $P=21.8$ day will be a strong evidence for V 885 Per being CHT. Alternatively, astrometric orbit with large period will support hierarchical quadruple configuration.
\par
Inner contact binary in V 885 Per shows time variable O'Connel effect, which indicates strong magnetic activity. It is good target for detailed studies of these phenomena.

\begin{acknowledgments}
 I am grateful for anonymous referee for a great suggestion to use {\sc KOREL}. I thank Guobao Zang for help with AI recovery of NSVS data.
Guoshoujing Telescope (LAMOST \cite{2012RAA....12.1197C, 2012RAA....12..723Z}) is a National Major Scientific Project built by the Chinese Academy of Sciences. Funding for the project has been provided by the National Development and Reform Commission. 
LAMOST is operated and managed by the National Astronomical Observatories, Chinese Academy of Sciences. The author gratefully acknowledge the ``PHOENIX Supercomputing Platform" jointly operated by the Binary Population Synthesis Group and the Stellar Astrophysics Group at Yunnan Observatories, Chinese Academy of Sciences. 
This work is supported by International Centre of Supernovae (ICESUN), Yunnan Key Laboratory of Supernova Research (No. 202505AV340004). 
\end{acknowledgments}

%
\vspace{5mm}

\section{DATA AVAILABILITY}
{\bf LAMOST spectra are available on \url{lamost.org/dr11}. Photometric datasets were downloaded from publicly available archives. Supporting materials like radial velocities and outputs for VO-KOREL runs including fitting logs have been uploaded to author webpage \url{mkuasbest.github.io/webpage/v885per.zip}, later will be also uploaded to zenodo}. 

\facilities{LAMOST, TESS, ASAS-SN, SuperWASP}





\appendix

\section{Spectrum taken close to periastron of SB1 orbit.}
\label{sec:peri}
\begin{figure*}
    \includegraphics[width=1\textwidth]{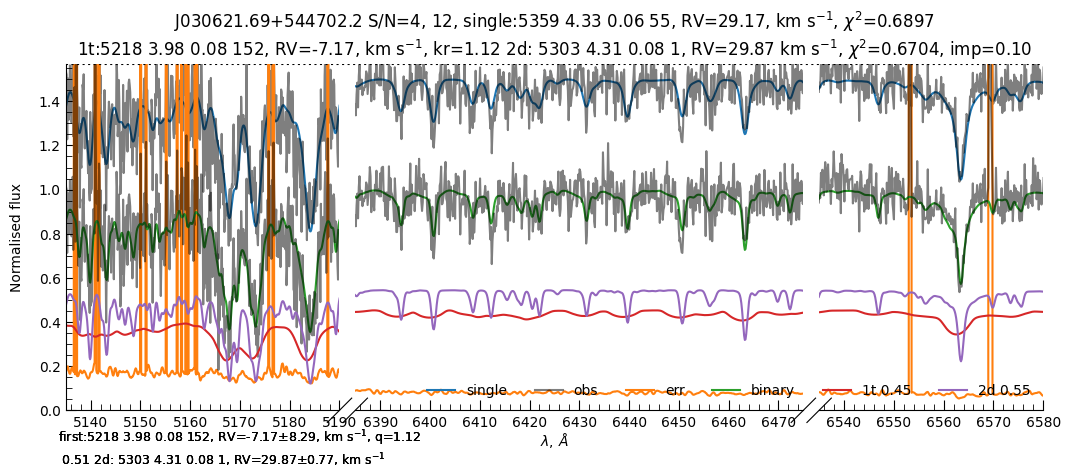}
     \caption{LAMOST MRS spectrum taken near periastron passage of star B, during the eclipse of V 885 Per. Title presents various technical information from the fit with single star and binary spectroscopic model. } 
    \label{fig:eclipse}
\end{figure*}

I show spectrum taken during eclipse of inner pair and close to periastron passage of star B in Figure~\ref{fig:eclipse}.
Stars in inner system have nearly identical radial velocities $\rv_1\sim\rv_2\sim\gamma_A=-7.17\pm8.29~\kms$ during the eclipse, which is in opposite phase relative Star B $\rv_B=29.87\pm0.77~\kms$. This qualitatively support connection between star B and inner system, although precision of $\gamma_A$ is too small to make robust conclusion.    

\section{recovery of NSVS LC and W-D estimations of minima times.}
\label{sec:ai}

\begin{figure}
    \centering
    \includegraphics[width=0.9\linewidth]{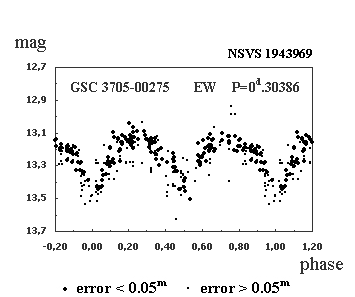}
    \includegraphics[width=0.9\linewidth]{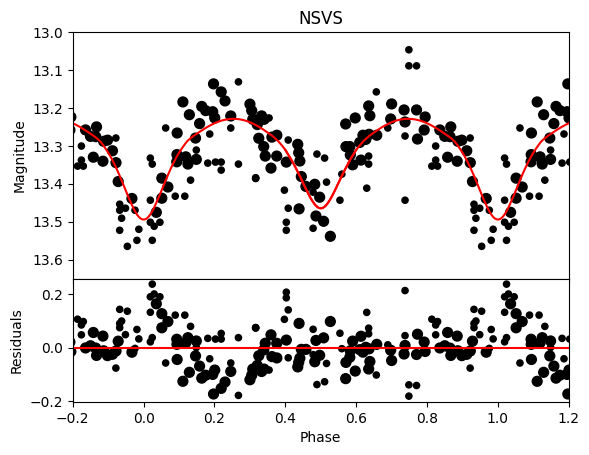}
    \caption{Original V 885 Per discovery plot from \protect\cite{2007PZP.....7....6K} (top panel) and my recovery of LC and best-fit model by W-D (bottom panel). Note that phase in bottom panels was added 0.5 for compatibility with top panel: dipper minimum of W-D model should have phase 0.5. }
    \label{fig:nsvs}
\end{figure}

Since NSVS LC is not available anymore, I tried to recover it using discovery paper \cite{2007PZP.....7....6K}, which presents phased plot (simple image with size 355x297 pixels available at \url{https://www.astronet.ru/db/varstars/msg/eid/PZP-07-0006} under index 5) with photometry shown as small and big circles, based on uncertainty, see top panel of Figure~\ref{fig:nsvs}. First I used ChatGPT to extract datapoints from this image, which worked surprisingly well, except for crowded regions, where it usually extracted only one datapoint. I got LC with 108 datapoints, which is significantly smaller in original LC (281 ``good" points in Skydot catalog available on VizieR). Therefore it is reasonable to assume that crowded regions should contain more points and many of them overlap with each other, so I manually added 33 missing datapoints. Weights for manually added datapoints were set six times larger in comparison with others. Times for each datapoint were calculated using {\bf ephemeris} from \cite{2007PZP.....7....6K} and this LC was normalised and fed to W-D code, where I assumed $R$ photometric band, following \cite{nsvs}. Fitting parameters for differential correction program include $t_0, L_1$ and $L_3$. Period was fixed to value from \cite{2007PZP.....7....6K}, with other parameters taken from ASAS-SN $g$ fit. Results are collected in Table~\ref{tab:wdtess}. New $t_0=51504.606\pm0.001$ day agrees with value from \cite{2007PZP.....7....6K} ephemeris: $t_0=51504.606$ day which I added half of their period for compatibility, since this paper reported equal minima min=13.45 mag.    

\begin{figure}
    \centering
    \gridline{\fig{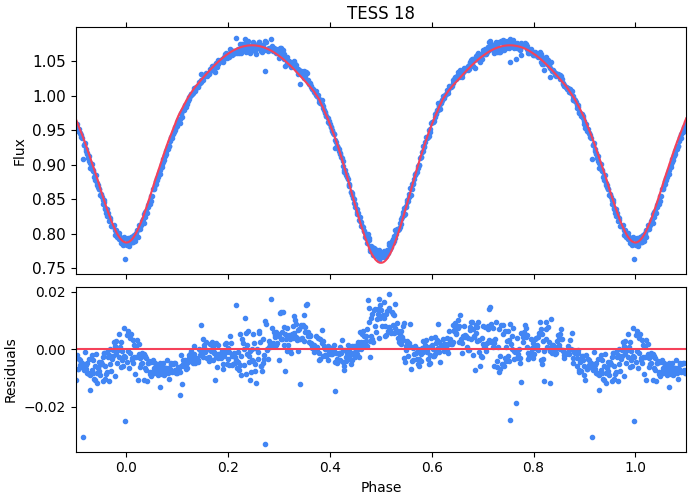}{0.95\columnwidth}{}}
    \gridline{\fig{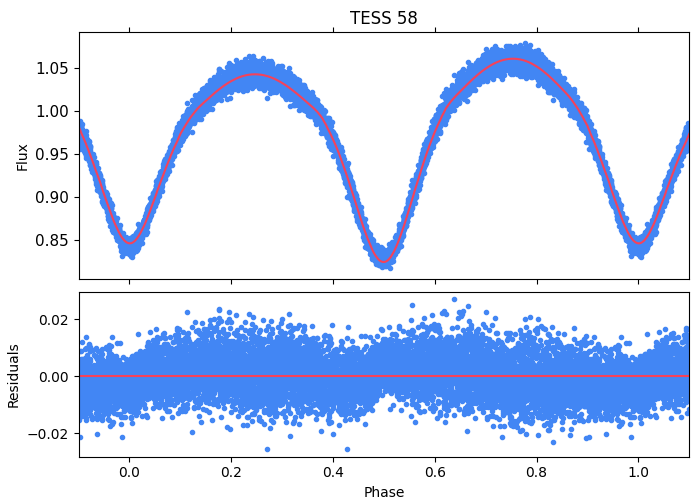}{0.95\columnwidth}{}}
    \gridline{\fig{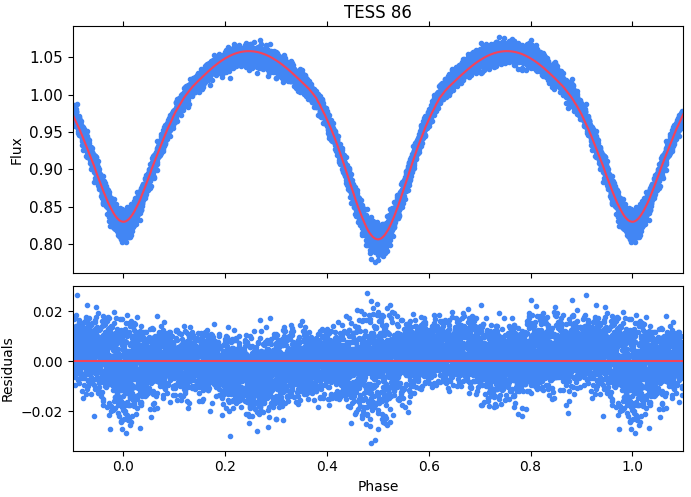}{0.95\columnwidth}{}}
    \caption{TESS data modeled with {\sc W-D}.}
    \label{fig:tess}
\end{figure}

For other LC datasets, which contain enough datapoints to cover full phase range in time interval shorter than one year, were used W-D to measure $t_0$. I use same parameters as for ASAS-SN $g$ LC fit, except for $L_1$ and $L_3$ because they are passband specific and can be affected by data reduction \citep{rgb2}. Also I fit for period $P_A$ in TESS and SuperWASP LCs since large number of datapoints allows for it. LC downloaded from O-C gateway were observed by Snaevarr Gudmundsson and Martin Lehky during one or two consecutive nights. For SuperWASP LC I selected three subsets observed by different cameras. The O'Connel effect is clearly visible in some datasets (i.e. TESS 58), therefore I added cool spot on the cooler primary component with manually pre-adjusted spot parameters.
All parameters are listed in Table~\ref{tab:wdtess}. Best fit models for TESS LCs are shown in Figure~\ref{fig:tess}. I stress that the detailed modeling of LCs is not the main topic of this paper, these W-D models are simple approximations of LC which can be used later as templates for measuring LTTE.

\begin{table*}[]
    \centering
    \begin{tabular}{c|ccccc}
    \hline
      LC   & $t_0$, day &$P_A$, day  & $L_1$ & $L_3$ \% & Spot ($\lambda.\phi,T_{\rm spot}/T_{\rm star},R/R_\ast$)\\
      \hline
NSVS (recovery)& 51504.606$\pm$0.001 & 0.30386(*) & 4.05$\pm$0.25 & 52.5$\pm$3.1&\\
SuperWASP 148 & 54395.9263$\pm$0.0004 &0.30387& 4.61$\pm$0.08& 40.3$\pm$1.0 & \\
SuperWASP 145 &54395.9259$\pm$0.0002 & 0.303871$\pm$0.000001& 4.34$\pm$0.03& 42.7$\pm$0.4& \\
SuperWASP 103 &53200.1988$\pm$0.0007 & 0.30387 & 4.17$\pm$0.19 & 44.6$\pm$2.5 & \\
TESS 86 & 60646.54623$\pm$0.00003 & 0.303869$\pm$0.000001 & 4.382$\pm$0.005 & 42.6$\pm$0.1& \\
TESS 58 & 59889.90754$\pm$0.00003 & 0.303869$\pm$0.000001 & 4.052$\pm$0.004 & 47.3$\pm$0.1 & $270^\circ,90^\circ,0.9,0.3$\\
TESS 18 & 58808.12812$\pm$0.00006 & 0.303873$\pm$0.000002 & 5.50$\pm$0.01& 28.0$\pm$0.1& \\
ZTF $r$ (**) & 58833.0452$\pm$0.0003 & 0.30387 & 5.62$\pm$0.10 & 26.3$\pm$1.3 & \\
Gudmundsson $R$ & 59213.7958$\pm$0.0001 & 0.30387 & 5.23$\pm$0.02 & 32.5$\pm$0.3 &  $327^\circ,70^\circ,0.9,0.3$\\
Gudmundsson $R$ &59224.7350$\pm$0.0001 & 0.30387 & 5.25$\pm$0.03 & 30.3$\pm$0.3 & $90^\circ,90^\circ,0.9,0.35$\\
Lehky $R$& 56949.0460$\pm$0.0007 & 0.30387 & 5.90$\pm$0.18 & 22.5$\pm$2.3 &$90^\circ,90^\circ,0.9,0.4$\\
Lehky $R$ & 56175.0884$\pm$0.0003 & 0.30387 & 5.55$\pm$0.07 & 29.3$\pm$0.9& $258^\circ,90^\circ,0.9,0.35$\\
Lehky $R$ & 55835.0563$\pm$0.0002 & 0.30387 &5.94$\pm$0.05 &24.9$\pm$0.6& $258^\circ,90^\circ,0.9,0.35$\\
Lehky $R$ & 57297.8900$\pm$0.0002 & 0.30387 &5.08$\pm$0.05 &32.2$\pm$0.6& \\
Lehky $R$ & 56642.7449$\pm$0.0002 & 0.30387 &6.11$\pm$0.07 &23.1$\pm$0.8& $229^\circ,109^\circ,0.9,0.35$\\
ASAS-SN $g$(***) & 59496.3955$\pm$0.0003& 0.30387 & 4.94$\pm$0.08 &33.4$\pm$1.0  &\\
ASAS-SN $g$(***) &60194.3861$\pm$0.0003& 0.30387 & 4.85$\pm$0.08 &34.9$\pm$1.0 &\\
ASAS-SN $g$(***) &60960.4446$\pm$0.0003& 0.30387 & 5.03$\pm$0.08 &32.8$\pm$1.0 &\\
ASAS-SN $g$(***) &58505.4720$\pm$0.0004& 0.30387 & 4.98$\pm$0.09 & 32.7$\pm$1.1 &\\
ASAS-SN $V$(***) &57982.5113$\pm$0.0003& 0.30387& 5.11$\pm$0.09 &32.9$\pm$1.1 &\\
      \hline
    \end{tabular}
    \caption{Best fit parameters for LC datasets using W-D. (*)- fixed to value from \protect\cite{2007PZP.....7....6K}. (**) 10 day long dense subset of original LC. (***) 200 day long subset of LC.}
    \label{tab:wdtess}
\end{table*}

\section{Simulation of CHT induced LTTE in TESS data.}
\label{sec:tess}

\begin{figure*}
    \centering
    \rotatefig{90}{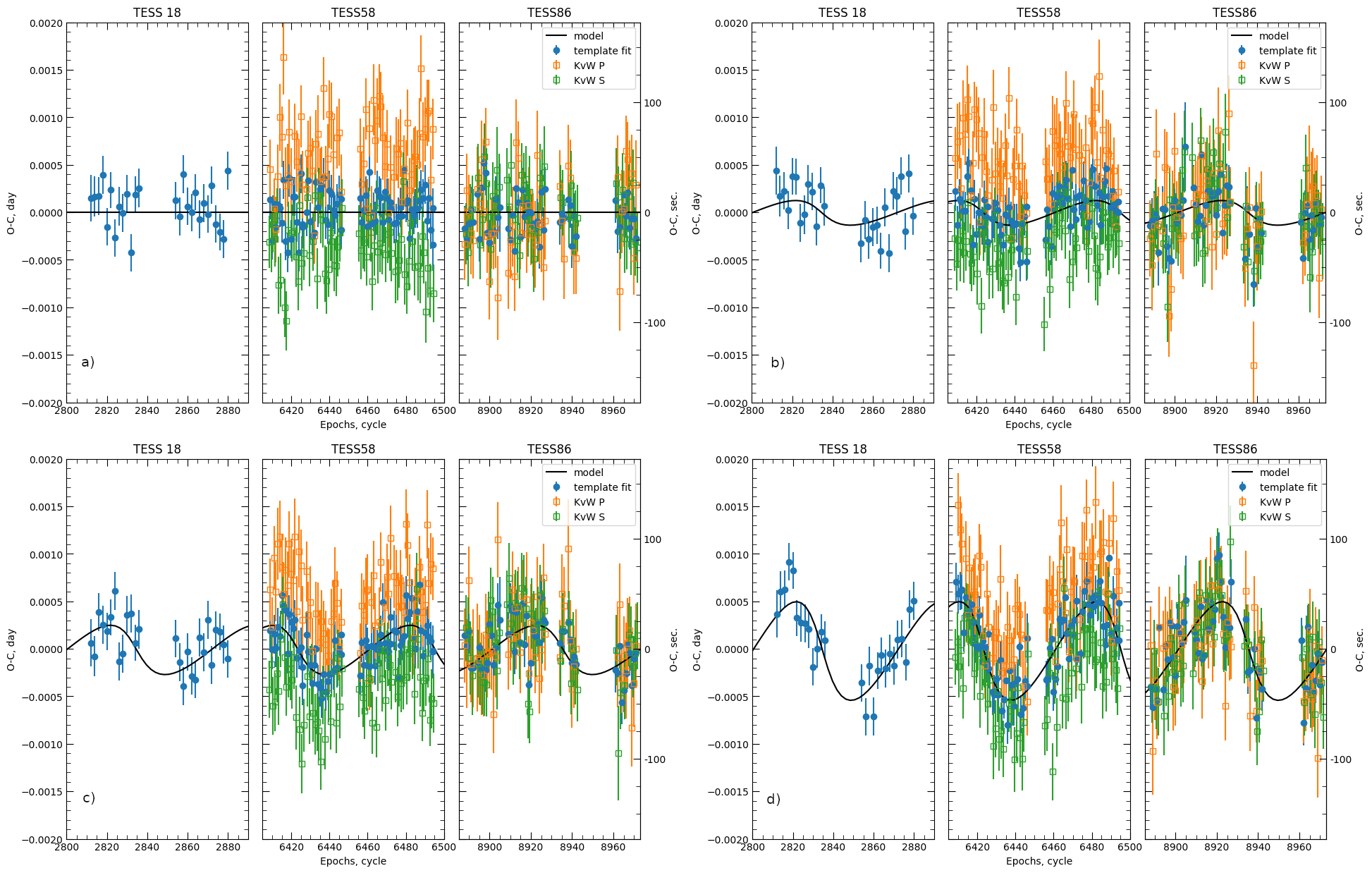}{1.15\textwidth}{}
    \caption{Same as Figure~\ref{fig:kvw}, but with analysis of simulated TESS data assuming different mass ratios: 0.0 (panel a), 0.5 (panel b), 1.0 (panel c), 2.0 (panel d). }
    \label{fig:tess_sym}
\end{figure*}

I used best fit TESS W-D models to compute synthetic LC for each TESS dataset, where I artificially include LTTE caused by Star B assuming different mass ratios $M_B/M_A=0.0,~0.5,~1.,~2.0$. Zero mass ratio case means that Star B has no effect on inner eclipsing system, this set provides proxy to efficiency of ETV derivation method. I measured ETV from these synthetic LC using same template fitting and KvW methods, same way as from real observational datasets, see Figure~\ref{fig:tess_sym}. For big mass ratios 1.0 and 2.0 LTTE has large amplitude $\mathcal{A_{\rm LTTE}}_A\sim22, 44$ s respectively and it can be clearly seen in ETV curves, even if each measurement has relatively large error $\sim40$ s. Even TESS 18 LC with long cadence of 1800 s allows to detect it is using template fitting technique. However Star B is likely less massive than inner contact system, so $\mathcal{A_{\rm LTTE}}_A\leq22$ s. Panel b) shows this example and it is very hard to reliably see LTTE here, therefore no detection of LTTE in real TESS data can't reject possibility that Star B is bound to inner contact system.   

\bibliographystyle{aasjournal}


\end{document}